\documentstyle[preprint,aps,eqsecnum]{revtex}

\begin{document}
\draft
\title
{Bosonization of Fermi Liquids}
\author{A.~H.~Castro Neto and Eduardo Fradkin}
\bigskip
%
\address
{Loomis Laboratory of Physics\\
University of Illinois at Urbana-Champaign\\
1100 W.Green St., Urbana, IL, 61801-3080}
%
\maketitle

\begin{abstract}
We bosonize a Fermi liquid in any number of dimensions in the limit
of long wavelengths. From the bosons
we construct a set of coherent
states which are related with the displacement of the Fermi surface
due to particle-hole excitations. We show that an interacting hamiltonian
in terms of the original fermions is quadratic in the bosons.
We obtain a path integral representation for the generating functional
which in real time, in the semiclassical limit, gives the Landau
equation for sound waves and in the imaginary time gives us the
correct form of the specific heat for a Fermi liquid even with the
corrections due to the interactions between the fermions. We also
discuss the similarities between our results and the physics of
quantum crystals.
\end{abstract}

\bigskip


\narrowtext

\section{Introduction}

Attempts to describe fermionic systems by bosons date to the
early days of second quantization.
In the early 50's Tomonaga ~\cite{tomonaga}, generalizing earlier work
by Bloch~\cite{bloch} on sound waves in dense Fermi systems,  gave an
explicit
construction of the Bloch waves for systems in one
dimension. Three years after the seminal work of Tomonaga, Bohm and
Pines~\cite{bohm} showed that there was a natural connection
between
bosonization and the Random Phase Approximation (RPA). Subsequently
many authors~\cite{authors} derived an explicit
Fermi-Bose transmutation in one-dimensional systems.
These works uncovered deep connections in
relativistic field theories (both fermionic and bosonic) and
with condensed matter systems.

The success of the bosonization approach in one dimension is
related to phase space considerations. Even for non-interacting fermions,
two excitations with arbitrarily low energies moving in the same
direction move at the same speed (the Fermi velocity) and,
hence, are almost a bound state. Consequently, even
the weakest interactions can induce dramatic changes in the nature of
the low-lying states.
These phase space effects
become manifest even
in perturbation theory and result in the presence of both marginal and
marginally relevant operators. Bosonization yields a simple and very
general description of the possible fixed point structure of strongly
correlated systems: marginally relevant operators are responsible for
the gap generating instabilities while strictly marginal operators give
rise to anomalous dimensions. Quite generally, the fixed points are
determined by the marginal operators alone. In turn, the anomalous
dimensions produce non-Fermi liquid behavior.
Bosonization in one spacial dimension is the
statement that the charge and current densities for states restricted
to the vicinity of the two Fermi points, $\rho(x)$
and $j(x)$, of electrons obey the equal-time algebra (Kac-Moody)
\begin{equation}
[\rho(x),j(y)]=-\frac{i}{\pi}
{\frac{\partial}{\partial x}}\delta(x-y)
\end{equation}
This algebra implies that there
exists of a free bosonic field $\phi(x)$ and its canonically conjugate
momentum $\Pi(x)$ which obey the canonical equal-time commutation
relation
\begin{equation}
[\phi(x),\Pi(y)]=i \delta(x-y)
\end{equation}
provided that one makes the identifications
\begin{equation}
\rho(x) \equiv {\frac{1}{\sqrt \pi}}{\frac{\partial}{\partial
x}}\phi(x), \qquad j(x) \equiv {\frac{1}{\sqrt \pi}}\Pi(x)
\end{equation}
More complicated operators, such as various order parameters associated
with $2 k_F$ excitations, can also be identified with suitably chosen
bosonic operators \cite{mandelstam}. Typically, these operators involve
non-local or
non-linear (or both) expressions in the bosonic field. In any event, the
basic building blocks are the local density and current operators which
describe electron-hole pairs at small momentum transfers.

In dimensions
higher than one, phase space considerations change the physics of the
low-lying states. The only marginal operators left ({\it i.~e.~},
operators whose coupling constants (Landau parameters) do not change as
the energy scale is lowered) are responsible only
for changing the shape of the Fermi surface. However the low temperature
and low frequency behavior of physical quantities is insensitive to the
presence of such marginal operators. In a sense, these operators are
redundant.
This
observation is at the root of the stability of the Landau theory of the
Fermi liquid~\cite{pomeranchuk,shankar}. It is, thus, hardly surprising
that very few attempts have been made to generalize the bosonization
approach to dimensions higher than one. In addition, phase space
considerations also tell us that even for small momentum transfers, an
electron-hole pair can decay into its electron and hole constituents.
Electron-hole collective modes do exist ( and are described by the
Random Phase Approximation) but so does the electron hole continuum.
Hence, unlike what happens in one space dimension, kinematics alone does
not require the generic existence of a bosonic bound state.

The first serious
attempt at bosonization in higher dimensions was carried out by Luther
{}~\cite{luther} who constructed a generalized bosonization formula
in terms of the fluctuations of the Fermi sea along radial directions in
momentum space. However this approach was not pursued.
Interest in the construction of bosonized versions of Fermi liquids has
been revived recently in the context of strongly correlated
systems~\cite{anderson} due to the possibility of non-conventional ground
states which are not Fermi liquid like. Since in one dimension the
bosonization approach is a powerful non-perturbative tool, the
expectation is that if a suitably generalized analog does exist in
higher dimensions, it may throw light on the non-perturbative aspects of
Fermi liquids and, hopefully, it may help to define non-Fermi liquids.
It is well known since the early days of the Landau theory that particle-hole
excitations have bosonic character {\it e.~g.~},
the sound waves (zero sound collective modes) of the Fermi surface of
neutral liquids
or plasmons in charged Fermi liquids~\cite{bohm}.
Inspired by these identifications, and drawing from Luther's
work, Haldane~\cite{haldane} has derived recently a bosonic algebra
for density fluctuations of a Fermi liquid in the form of a generalized
Kac-Moody algebra which generalizes eq~(1.1).
Haldane's construction has been examined recently by several authors
{}~\cite{houghton,khveshchenko}.

Our approach has the same starting point as
Haldane's.
We also begin by constructing a bosonic algebra with creation and
annihilation
operators which act over a reference state describing a filled Fermi
sea. This algebra is valid in a restricted Hilbert space
of states generated by small deformation of the filled Fermi sea.
Using these operators we construct a set of coherent states
which can be regarded as deformation of
the Fermi surface.
More precisely,
we show that the eigenstate of the annihilation operators of the
algebra, namely
the coherent states, are related with the displacement of the Fermi
surface in some direction. These displacements are a coherent
superposition
of particle-hole excitations close to the Fermi surface.
This interpretation leads to the picture of
the Fermi surface as a real dynamical object. We have
two
strong arguments in support of this picture. The first argument is based
on the fact that, using these coherent states, we reproduce the
phenomenological theory of
Pomeranchuk~\cite{pomeranchuk} on the stability of a Fermi liquid and
with its
modern reinterpretation (by Shankar~\cite{shankar}) in terms of the
renormalization group method. Pomeranchuk's approach is based on
the work of Landau on many-body systems.
The second argument follows from the fact (which we elaborate further
below) that it is possible to define a surface
tension associated with the Fermi surface itself.

Armed with these coherent states, we construct a coherent state path
integral which
can be viewed as a sum over the histories of the shape of the Fermi
surface, a bosonic {\it shape field}. We show that there exists a class
of simple interacting hamiltonians (in terms of the original fermions
of the theory),
which are naturally related to the Landau theory of the Fermi liquid,
which in the coherent state bosonization leads to a simple quadratic
hamiltonian in terms of the bosons. These hamiltonians are the natural
generalization of the Luttinger/Thirring models of one dimensional
systems. According to the renormalization group analysis, these
hamiltonians represent the stable fixed points of Fermi liquids.

With the path integral we can study the thermodynamics and dynamics
of interacting electronic systems in a very simple way. In order
to prove the consistency of our method we calculate the low
temperature specific
heat and obtain the same results of the Fermi liquid theory (although
we have only {\it real} bosons in the problem). We can go even
further and calculate the corrections for the specific heat due
to interactions and we show that our results also agree with the
well known results of the Fermi liquid theory. This calculation
reveals the new features of the bosonic field involved in the
problem. They are not an usual free bosonic field but a {\it
topologically constrained excitation}. The constraint comes from the
need to sum over the bosonic excitations {\it tangent} to the Fermi
surface and do not contribute to the energetics. This is most important
since free non-relativistic bosons in space dimensions larger than one
do not have the linear specific heat which is characteristic of Fermi
liquids.

In this paper we discuss the bosonization approach to the fixed point
hamiltonians. These hamiltonians contain operators which are at most
marginal. We deliberately leave out a number of relevant operators
which lead to the well known instabilities of the Fermi liquid:
superconductivity, magnetism, etc. We will discuss these operators
elsewhere.

We also show
that the Landau theory is the semiclassical approximation
for the bosons which is exact for the fixed point hamiltonians since the
hamiltonian is quadratic in bosons.
This result means that the bosonic field is nothing but the field
of sound waves which propagates on the Fermi surface.
We applied these methods to study the physics of sound waves in two
dimensions.
We obtain a general
equation for the sound modes and, in particular, we study the
zero sound. In order to study the characteristics of these sound
modes we go to the limit of large Fermi momentum and concentrate in
the forward scattering direction. We introduce a new scale in the
problem, the range of interaction. We find a rather complex spectrum of
sound modes which describe increasingly wrinkled
Fermi surfaces. In the
case of long range interactions,
the shape of the Fermi surface can become unstable. The existence
of strong local quantum fluctuations of the shape of the Fermi surface
also suggests an analogy with the quantum mechanical fluctuations of the
shape of a crystal at zero temperature. Thus it is natural to ask
if it is possible that,
for long range interactions,
the Fermi surface could undergo a
roughening transition. In the case of
of the shape of a three dimensional crystal it is known ~\cite{fradkin}
that quantum
fluctuations generally make the surface smooth, not rough.
However, in the case of a ``planar crystal" , which has a one
dimensional surface, it is possible to have a quantum mechanical
roughening transition. It turns out, however, that by an explicit
calculation of the correlation function between different pieces of the
Fermi surface we can show that quantum fluctuations of the fixed
point hamiltonian wash out this interesting possibility. The reason is
that the dynamics described by the fixed point hamiltonians enforce
Luttinger's theorem as a local condition. It may be possible to find
other hamiltonians which enforce Luttinger's theorem only as a global
constraint. The shape of the Fermi surface of the ground states  of
these hamiltonians
may exhibit a quantum roughening transition in two space dimensions.

The paper is organized as follows. In sections II and III we present two
physical arguments which suggest that the Fermi surface should be
regarded as a quantum mechanical object. In section II we review the
stability of a Fermi liquid and specialize the discussion to the case of
two dimensions . In section
III we refine the analogy with the theory of surfaces by defining the
surface tension of the Fermi surface. In section
IV we describe the coherent state bosonization construction for an
interacting
fermionic system. In section V we show how a fermionic hamiltonian
can be rewritten in terms of the bosons and in section VI we study
the Fermi liquid properties, the thermodynamics and classical dynamics.
Section VII contains our conclusions.

\section{The stability of the Fermi liquid in two dimensions}

The route for the stability of a Fermi liquid was established by Pomeranchuk
{}~\cite{pomeranchuk} who studied the effect of the change of the Fermi surface
in the free energy of the system
using the Landau expansion for the change in the total energy,
\begin{equation}
\Delta E= \sum_{\vec{p}} \epsilon_{\vec{p}}^{0} \delta n_{\vec{p}}
+ \frac{1}{V}\sum_{\vec{p},\vec{p'}} f_{\vec{p},\vec{p'}} \delta n_{\vec{p}}
\delta n_{\vec{p'}},
\end{equation}
where $V$ is the volume of the system, $\delta n_{\vec{p}}$ is the
deviation of the occupation number
with $\vec{p}$ is at the Fermi surface,
$\epsilon_{\vec{p}}^{0}$ is the bare dispersion relation and
$f_{\vec{p},\vec{p'}}$ is the quasi-particle interaction.

In this section we essentially follow the work of Pomeranchuk
{}~\cite{pomeranchuk,baym} but
we restrict ourselves to two dimensions since in three dimensions
the results are well known and two-dimensional Fermi liquids are of
direct physical interest. Consider a change in the Fermi momentum given
by \begin{equation}
p_F(\theta) = p_F + \delta p_F(\theta)
\end{equation}
where $\theta$ is the angle which parametrizes the position of the
Fermi surface and $p_F$ is the original Fermi momentum.

The new occupation number is simply given by
\begin{equation}
n_{\vec{p}} = \Theta(p_F(\theta)-p),
\end{equation}
substituting (2.2) in (2.3) and expanding up to second order in the
deviations we find
\begin{equation}
\delta n_{\vec{p}} = \delta p_F(\theta) \delta(p_F-p) -
\frac{(\delta p_F(\theta))^2}{2} \frac{ d \delta(p_F-p)}{d p}.
\end{equation}

Now, using (2.1) and (2.4) it is easy to show that the total change in
the free energy, $ F = E - \mu N$ ($\mu$ is the chemical potential and
$N$ is the total number of particles )
is given by
\begin{equation}
\frac{\Delta F}{V} = \frac{p_F v_F}{2 \pi} \int_{0}^{2 \pi} \frac{d\theta}{2
\pi}
(\delta p_F(\theta))^2 + \frac{p_F^2}{2 \pi^2} \int_{0}^{2 \pi}
\frac{d\theta}{2 \pi} \int_{0}^{2 \pi} \frac{d\theta'}{2 \pi}
f(\vec{p},\vec{p'}) \delta p_F(\theta)\delta p_F(\theta'),
\end{equation}
where $\vec{p}$ and $\vec{p'}$ are at the Fermi surface and $v_F$
is the Fermi velocity.

We expand the interaction and the displacement of the Fermi
surface in Fourier components (in two dimensions we can parametrize
the points of the Fermi surface by one angle, $\theta$ ) as,
\begin{equation}
f(\vec{p},\vec{p'}) = \sum_{m=0}^{+\infty} f_m \cos(m(\theta-\theta'))
\end{equation}
and
\begin{equation}
\delta p_F(\theta) = \sum_{m=0}^{+\infty} u_m \cos(m \theta),
\end{equation}
substituting these expressions in (2.5) and using the orthogonality between the
Fourier components one finds,
\begin{equation}
\frac{\Delta F}{V} = \frac{p_F^2}{8 \pi^2} \sum_{m=0}^{+\infty}
\left(\frac{1}{N(0)} + f_m\right) u_m^2
\end{equation}
where $N(0) = \frac{p_F}{\pi v_F}$ is the density of states at the
Fermi surface.

Observe that the Fermi liquid is stable if $\Delta F \geq 0$ or
\begin{equation}
N(0) f_n \geq -1
\end{equation}
for all $n$. This result implies that the Fermi liquid is a local
minimum in the configuration space of the many-body system. Local
stability, in terms of renormalization group, means that the effective
hamiltonian is at a fixed point~\cite{shankar}.

We notice that it is not a coincidence that the free energy (2.8)
has the same form as the one for a drumhead. It means that the
Fermi surface is a dynamical object with elastic properties.
Furthermore, we see that the interaction term gives the same
order of contribution as the free term of the energy. In terms
of the renormalization group interpretation of the stability
of a Fermi liquid \cite{shankar} it means that the interaction
is a marginal operator while next order expansion in the functional
(2.1) will give only irrelevant operators which do not contribute
to the low energy physics of the problem.

Anderson~\cite{anderson} has argued that a singular interaction
between the fermions could give rise to a new features which could
explain the anomalous behavior of cuprates. Haldane~\cite{haldane}
has constructed an argument which is mainly based on the form of
expression (2.8) which we reproduce here due to its simplicity and
elegance. Suppose that instead of an overall density of states,
$N(0)$, we introduce a {\it local} density of states at each
point of the Fermi surface. In order to do so we have to define
a cut-off on the Fermi surface, $\Lambda$, which is much smaller
than the Fermi momentum but yet large enough to count a macroscopic
number of states. By naive dimensional analysis we conclude that
this local density of states must scale like
$\Lambda^{d-1} v_F^{-1}$.
Therefore, from (2.8), the quantity $F_m=\Lambda^{d-1} v_F^{-1} f_m$
is the dimensionless coupling constant. Suppose that the interaction,
$f_m$, and the Fermi velocity are well behaved (they do not diverge)
in the limit where $\Lambda \to 0$ (the scaling limit). It is easy to
see that the $F_m$ goes to zero for $d > 1$ and therefore only the
non-interacting term in present. This result implies that the
interaction is {\it marginally irrelevant}. However, for $d=1$,
$F_m$ will be always finite, that is, the interaction is
{\it marginally relevant} in one dimension. Therefore, even if
the interaction is not singular in the scaling limit it has
profound consequences in one dimension while it ``scales away"
in higher dimensions. However, we have the interesting possibility
that the interaction term $f_m$ or the Fermi velocity diverge in
the scaling limit. If this is so, the dimensionless coupling
constant can diverge and therefore the operator becomes {\it
relevant} with respect to the non-interacting term. However,
this singular behavior is never present in a Fermi liquid
since the renormalization group flow is dominated by the
non-interacting fixed point.

The approach in this
section is essentially phenomelogical because we have postulated
(2.1) in order to get the stability condition (2.9). In the next
section we study the elastic properties of the Fermi surface from
the microscopic point of view. This new point o view will bring
a different perspective on the same subject, namely, the Fermi
surface as a {\it real quantum object}.

\section{The tension of the Fermi surface}

In this section we explore in more detail the picture of the Fermi
surface as a quantum object. In particular we will establish that, for
systems which obey the hypothesis of a Fermi liquid, it is possible to
define a surface tension. Here we follow the work
of Luttinger \cite{luttinger} on the properties of many-body systems.
It can be shown that the total energy of a spinless interacting
fermionic system can be written in the
form \cite{kadanoff} of an integral over the whole momentum space as,
\begin{equation}
E=\sum_{\vec{k}} E_{\vec{k}}
\end{equation}
where
\begin{equation}
E_{\vec{k}}= \int_{-\infty}^{+\infty} \frac{dw}{2\pi i} \, \, e^{iw\eta}
\, \, \frac{\left(w+\epsilon_{\vec{k}}^{0}\right)}{2} \, \, G(\vec{k},w)
\end{equation}
where $\eta \to 0^{+}$, $\epsilon_{\vec{k}}^{0}$ is the non-interacting
dispersion relation, $G(\vec{k},w)$ is the interacting Green's function
in momentum space.

Equation (3.2) can be rewritten in terms of the spectral function of
the system as \cite{nozieres},
\begin{equation}
E_{\vec{k}}= \int_{-\infty}^{+\mu} \frac{dw}{\pi} \, \,
\frac{\left(w+\epsilon_{\vec{k}}^{0}\right)}{2} \, \, A(\vec{k},w)
\end{equation}
where the spectral function $A(\vec{k},w)$ is given by
\begin{equation}
A(\vec{k},w)= \frac{\Sigma_I(\vec{k},w)}{
\left(w-\epsilon_{\vec{k}}^{0}-\Sigma_R(\vec{k},w)\right)^2 +
\Sigma_I(\vec{k},w)^2}
\end{equation}
where $\Sigma_R(\vec{k},w)$ and $\Sigma_I(\vec{k},w)$ are the
real and imaginary part of the self energy and $\mu$ is the
chemical potential of the problem.

The spectrum of the interacting system (the poles of the Green's
function), $\epsilon_{\vec{k}}$,
is given by the solution of the self consistent equation
\begin{equation}
\epsilon_{\vec{k}} - \epsilon_{\vec{k}}^{0} -
\Sigma_R(\vec{k},\epsilon_{\vec{k}})=0
\end{equation}
and, in particular, the Fermi surface is defined by the set
of vectors $\{\vec{p}_F\}$ such that
\begin{equation}
\epsilon_{\vec{p}_F}=\mu.
\end{equation}

It is clear from the above  definitions that we expect that any
singularity in this problem should appear near the Fermi surface. We,
therefore, split the integral in (3.3) in two pieces, from $-\infty$
to $\mu - \zeta$ (free of singular terms) and from $\mu - \zeta$ to
$\mu$ (the singular contribution) where $\zeta$ is
an small energy scale which permit us to look only to the singular
part of $E_{\vec{k}}$.

Moreover, consider the vectors $\vec{k}$ close to the Fermi surface,
\begin{equation}
\vec{k}=\vec{p}_F+\delta\vec{k}
\end{equation}
where the displacement $\delta\vec{k}$ is perpendicular to
the Fermi surface, that is,
\begin{equation}
\delta \vec{k}.\nabla \epsilon_{\vec{p}_F}= \delta k \mid \nabla
\epsilon_{\vec{p}_F} \mid
\end{equation}
and $\delta k$ is positive if the vector points outside of the
Fermi surface and negative it points inside the Fermi surface.

Since we are interested in the integral close to the Fermi surface
we can expand the integrand using the above formulae. In particular
the denominator in (3.4) can be rewritten as
\begin{equation}
w-\epsilon_{\vec{k}}^{0}-\Sigma_R(\vec{k},w)= Z_{\vec{k}}^{-1}
(w-\epsilon_{\vec{k}})
\end{equation}
where
\begin{equation}
Z_{\vec{k}}^{-1} = 1 - \left(\frac{\partial\Sigma_R(\vec{k},w)}
{\partial w}\right)_{w=\epsilon_{\vec{k}}}
\end{equation}
is the quasi-particle residue~\cite{nozieres}.

We can also write
\begin{equation}
\epsilon_{\vec{k}}= \mu+ \delta k \mid\nabla\epsilon_{\vec{p}_F}\mid
\end{equation}
where we used (3.6), (3.7) and (3.8).

Substituting these approximations in the singular part of the integral
(3.3) and changing variables with respect to the chemical potential we get
\begin{equation}
E^{s}_{\vec{k}}=\int_{0}^{\zeta} \frac{dw}{\pi} \, \,
\frac{\left(\mu+\epsilon^{0}_{\vec{k}}-w\right)}{2} \, \,
\frac{\Sigma_I(\vec{k},\mu-w)}{\left(Z_{\vec{k}}^{-2}
\left(w+\delta k \mid \nabla \epsilon_{\vec{p}_F}\mid\right)^2 +
\Sigma_I(\vec{k},\mu-w)^2\right)}
\end{equation}

For a Fermi Liquid we use the fact that the imaginary part of the self
energy can be written as \cite{luttinger}-\cite{nozieres},
\begin{equation}
\Sigma_I(\vec{k},w)= C_{\vec{k}} (w-\mu)^2 sgn(\mu-w)
\end{equation}
where $C_{\vec{k}}$ depends only on $\vec{k}$ and $sgn(x) = 1(-1)$ if
$x>0(x<0)$.

Substituting (3.13) in (3.12), changing the variables of integration
from $w$ to $x=\frac{\mid\delta k\mid \mid \nabla \epsilon_{\vec{p}_F}\mid}{w}$
we found,
\begin{equation}
E^{s}_{\vec{k}}=\int^{\infty}_{\frac{\mid\delta k\mid \mid \nabla
\epsilon_{\vec{p}_F}\mid}{\zeta}} \frac{dx}{2 \pi} \,
\left(\mu+\epsilon^{0}_{\vec{k}}-\frac{1}{x} \mid\delta k\mid
\mid \nabla \epsilon_{\vec{p}_F}\mid\right)
\frac{C_{\vec{k}} \mid\delta k\mid \mid \nabla \epsilon_{\vec{p}_F}\mid}
{\left(Z_{\vec{k}}^{-2} x^2 (x+sgn(\delta k))^2 + C_{\vec{k}}^2 \, \delta k^2
\mid \nabla \epsilon_{\vec{p}_F}\mid^2\right)}.
\end{equation}

Observe now that if we let $\delta k \to 0$ we can use the well known
expression
\begin{equation}
Lim_{\mid \delta k \mid \to 0} \frac{C_{\vec{k}} \mid\delta k\mid \mid \nabla
\epsilon_{\vec{p}_F}\mid}
{\left(Z_{\vec{k}}^{-2} x^2 (x+sgn(\delta k))^2 +
C_{\vec{k}}^2 \delta k^2 \mid \nabla \epsilon_{\vec{p}_F}\mid^2\right)} =
\pi \mid Z_{\vec{p}_F} \mid
\left( \delta(x+sgn(\delta k)) - \delta(x) \right).
\end{equation}
Substituting this result in (3.14) we find
\begin{equation}
E^{s}_{\vec{k}}=\mid Z_{\vec{p}_F}\mid \, \,
\frac{\left(\epsilon^{0}_{\vec{p}_F}+\mu\right)}{2} \Theta(-\delta k)
\end{equation}
where $\Theta(x)=1 (0)$ if $x>0 (x<0)$.

{}From (3.16) we conclude therefore that there is a discontinuity at the Fermi
surface which is given by
\begin{equation}
\Delta E(\vec{p}_F) = \mid Z_{\vec{p}_F}\mid \, \,
\frac{\left(\epsilon^{0}_{\vec{p}_F}+\mu\right)}{2}
\end{equation}

Therefore there is no singularity when the quasi-particle residue
vanishes and we have non-Fermi liquid behavior.
In the non-interacting system is easy to see that this discontinuity is
simply given by $\mu$ which is the energy needed to put another electron
in the system.

We can also show (see appendix A) that this result holds true if the
imaginary part of
the self energy goes like $(w-\mu)^{1+\nu}$ with $\nu$ greater than
zero. However, if $\nu$ is smaller than zero we can prove that the
discontinuity in the $E(\vec{k})$ vanishes as $\delta k^{(1-\nu)^2}$
as $\delta k \to 0$. Therefore, $\nu = 0$ is a critical point (this
is the case of a marginal Fermi liquid phenomenology~\cite{varma}).

We can interpret this discontinuity in $E(\vec{k})$ as due to a
surface tension in momentum space since $E(\vec{k})$ is the contribution
of the mode $\vec{k}$ to the total energy, as we see by (3.1).
The surface tension in three dimensions can be defined by the difference
of energy density across the Fermi surface as follows \cite{landau},
\begin{equation}
\sigma_F \left(\frac{1}{R^{1}_{F}} + \frac{1}{R^{2}_{F}}\right) =
\frac{\Delta E(\vec{p}_F)}{V_F}
\end{equation}
where $R^{1}_{F}$ and $R^{2}_{F}$ are the principal radii of
curvature at $\vec{p}_F$ and $V_F$ is the volume of the Fermi sea.

In two dimensions the analogous of (3.18) is
\begin{equation}
\frac{\sigma_F }{R_{F}} =\frac{\Delta E(\vec{p}_F)}{A_F}
\end{equation}
where $A_F$ is the area of the Fermi sea.

Comparing (3.18) or (3.19) with (3.17) we conclude that the
surface tension is proportional to the quasiparticle residue.
Therefore, for a non-Fermi liquid behavior the surface tension
vanishes with the
quasiparticle residue. From this observation we can conclude that
a transition from Fermi liquid to non-Fermi liquid behavior can
be viewed as a phase transition in momentum space for the case
where the quasiparticle residue can be tuned to zero adiabaticlly.

\section{Bosonization and coherent states}

Our starting point to approach the bosonization resembles the
microscopic approaches for the foundations of Fermi liquid
theory ~\cite{baym}. However, instead of working with the
dynamics of the response functions, we will work directly with the
properties of the operators which generate the physicals spectrum,
in a restricted Hilbert space of states of the filled Fermi sea. This is
the standard
procedure of bosonization in one dimension and in a recent work of
Haldane~\cite{haldane}. From this perspective, the algebra obeyed by the
operators is not a property of the operators themselves but a property
of the states.

For simplicity we will consider a system of interacting
spinless fermions. Generalizations to systems with an internal symmetry,
such as spin SU(2), can be done with some minor but important
modifications.
The density of fermions at some point
$\vec{r}$ at some time $t$ is given by
\begin{eqnarray}
\rho(\vec{r},t) = \psi^{\dag}(\vec{r},t) \psi(\vec{r},t)
= \sum_{\vec{k},\vec{q}} c^{\dag}_{\vec{k}-\frac{\vec{q}}{2}}(t)
c_{\vec{k}+\frac{\vec{q}}{2}}(t) \, e^{i \vec{q}.\vec{r}}
\end{eqnarray}
where $c^{\dag}_{\vec{k}}$ and $c_{\vec{k}}$ are the creation and
annihilation operator of an electron at some momentum $\vec{k}$
which obey the fermionic algebra,
$\left\{c^{\dag}_{\vec{k}},c_{\vec{k'}}\right\} =
\delta_{\vec{k},\vec{k'}}$,
where $\{...\}$ is the anticommutator and all other anticommutation
relations are zero.
In the Fermi liquid theory
{}~\cite{baym} the operator which
appears in the r.h.s. of (4.1) determines the behavior of the
system.
We will concentrate on this operator which we will denote by
\begin{equation}
n_{\vec{q}}(\vec{k},t)= c^{\dag}_{\vec{k}-\frac{\vec{q}}{2}}(t)
c_{\vec{k}+\frac{\vec{q}}{2}}(t).
\end{equation}
In particular, $n_{0}(\vec{k})$ is the number operator in momentum
space.
As in the standard approaches for the Fermi liquid theory we
will concentrate our interest in regions close to the Fermi surface
($\vec{k} \sim \vec{p}_F$) and consider long wavelength fluctuations
around these regions ($\vec{q} \to 0$).

The equal time commutation relation between the operators defined in (2)
is easily obtained. In the long wavelength limit ($ q << \Lambda$,
where $\Lambda$ is a cut off) we get
\begin{equation}
\left[n_{\vec{q}}(\vec{k}),n_{\vec{-q'}}(\vec{k'})\right]=
- \delta_{\vec{k},\vec{k'}} \delta_{\vec{q},\vec{q'}} \vec{q}.\nabla n_{0}
(\vec{k}) + 2 n_{0}(\vec{k}) \delta_{\vec{q},\vec{-q'}} \vec{q}.
\nabla \delta_{\vec{k},\vec{k'}}.
\end{equation}

We are interested in the behavior of systems with a Hilbert space
restricted to the vicinity of the Fermi surface. We define the Hilbert
space to be the filled Fermi sea $|FS >$ and
the tower
of states obtained by acting finitely on it with local fermion
operators. More specifically we consider a shell of states of thickness
$D$ (measured in units of momentum) around the Fermi energy. Next we
imagine dividing up the Fermi surface in patches~\cite{haldane,houghton}
centered at points $\vec k$ on the Fermi surface of the filled Fermi
sea. Each patch has thickness $D$ and width $\Lambda$ ( in momentum
units and tangent to the surface). Instead of the sharp operators
$n_{\vec{q}}(\vec{k})$ we will considered operators {\it smeared} over
each patch. We will replace eq~(4.3) by
a weaker identity valid in the restricted Hilbert space. Furthermore, we
will
make the explicit assumption that we are both in the thermodynamic limit
(the momenta form a continuum) and that the Fermi surface is {\it
macroscopically large} ($q \ll D \ll \Lambda \ll p_F$).

Since the vectors $\vec{k}$
are at the Fermi surface and $\vec{q}$ are very small,
in the restricted Hilbert space it is possible to
replace the  r.~h.~s.
of (4.3) by its expectation value in the filled Fermi sea $|FS >$ , namely,
\begin{eqnarray}
n_{0}(\vec{k}) \rightarrow \langle n_{0}(\vec{k}) \rangle =
\Theta(\mu-\epsilon_{\vec{k}})
\nonumber \\
\nabla n_{0}(\vec{k}) \rightarrow \nabla \langle n_{0}(\vec{k}) \rangle =
- { \vec{v}}_{\vec{k}} \delta(\mu-\epsilon_{\vec{k}})
\end{eqnarray}
where $\mu$ is the chemical potential, $\epsilon_{\vec{k}}$ is the
one-particle fermion spectrum (from which the Hilbert space is
constructed) and $\vec{v}_{\vec{k}} = \nabla \epsilon_{\vec{k}}$
the velocity of the excitations.
The terms ignored in (3) vanish as $\frac{\Lambda}{p_F^2}$  as the size
of
the Fermi surface diverges. More generally, all corrections vanish
if the limit $q \ll D \ll \Lambda \ll p_F$ is satisfied. These are
exactly the same assumptions
that enter in the construction in one-dimension (in appendix B we prove
the equivalence of our construction and the well known bosonization
in one dimension). Notice that the state $|FS
>$, which is used to normal-order the operators, is not necessarily the
ground state of the system of interest (as in the one-dimensional case).
However, this approach will succeed only if the true ground state belongs
to the restricted space.

Hence, within this approximation the commutators of the operators
$n_{\vec{q}}(\vec{k})$ become c-numbers, namely,
\begin{equation}
\left[n_{\vec{q}}(\vec{k}),n_{\vec{-q'}}(\vec{k'})\right]=
\delta_{\vec{k},\vec{k'}} \delta_{\vec{q},\vec{q'}} \vec{q}.\vec{v}_{\vec{k}}
\delta(\mu-\epsilon_{\vec{k}}).
\end{equation}
where we drop the last term in the r.h.s. of (4.3) because its matrix
elements near the Fermi surface are down by powers of $\frac{\Lambda}{p_F}$.
We now define the operator
\begin{equation}
a_{\vec{q}}(\vec{k}) = n_{\vec{q}}(\vec{k})
\Theta(sgn(q)) + n_{-\vec{q}}(\vec{k})\Theta(-sgn(q))
\end{equation}
where the $sgn(q)$ is $+1$ if $\vec{q}$ points outside the Fermi surface and
$-1$ if it points inside the Fermi surface at the point $\vec{k}$.
The adjoint of (4.6) is simply,
\begin{equation}
a^{\dag}_{\vec{q}}(\vec{k}) =
n_{-\vec{q}}(\vec{k}) \Theta(sgn(q)) + n_{\vec{q}}(\vec{k})\Theta(-sgn(q))
\end{equation}
where we used, from the definition (4.2), that
$n^{\dag}_{\vec{q}}(\vec{k}) = n_{-\vec{q}}(\vec{k})$.

It is important to notice that, by construction, the operator defined
in (4.6) annihilates the filled Fermi sea
\begin{equation}
a_{\vec{q}}(\vec{k}) \mid FS \rangle = 0.
\end{equation}
Moreover, from the commutation relations (4.5) we easily obtain,
\begin{equation}
\left[a_{\vec{q}}(\vec{k}),a^{\dag}_{\vec{q'}}(\vec{k'})\right]=
\mid \vec{q}.\vec{v}_{\vec{k}} \mid  \delta(\mu-\epsilon_{\vec{k}})
\delta_{\vec{k},\vec{k'}} \left(\delta_{\vec{q},\vec{q'}}
+\delta_{\vec{q},-\vec{q'}}\right),
\end{equation}
all other commutators vanish.

Eq. (4.8) and (4.9) show that the elementary excitations have bosonic
character,
particle-hole pairs, and they are created and annihilated close to the
Fermi surface by these operators, moreover, they span the Hilbert
space of low energy. But we need an interpretation for these excitations.
It is natural now to define the coherent state~\cite{stone}
\begin{equation}
\mid u_{\vec{q}}(\vec{k}) \rangle = U (\vec{k}) \mid FS \rangle
\end{equation}
where
\begin{equation}
U(\vec{k}) = \exp\left(\sum_{\vec{q}} \frac{v_{\vec{k}}}
{2 \mid \vec{q}.\vec{v}_{\vec{k}} \mid} \, u_{\vec{q}}(\vec{k}) \,
a^{\dag}_{\vec{q}}(\vec{k})\right)
\end{equation}
where the sum (and all other sums that follow are restricted to
$q << \Lambda$). Observe that from the definition (4.7) we have
$a^{\dag}_{-\vec{q}}
(\vec{k}) = a^{\dag}_{\vec{q}}(\vec{k})$ and we choose
$u_{\vec{q}}(\vec{k})=u_{-\vec{q}}(\vec{k})$. Using this property and
the commutation relation (4.9) we find
\begin{equation}
U^{-1}(\vec{k}) \, a_{\vec{q}}(\vec{k}) \, U(\vec{k}) =
a_{\vec{q}}(\vec{k}) + \delta(\mu-\epsilon_{\vec{k}}) v_{\vec{k}}
u_{\vec{q}}(\vec{k})
\end{equation}
which, together with (4.8), leads us to the eigenvalue equation,
\begin{equation}
a_{\vec{q}}(\vec{k}) \mid u_{\vec{q}}(\vec{k}) \rangle =
\delta(\mu-\epsilon_{\vec{k}}) \, v_{\vec{k}} \, u_{\vec{q}}(\vec{k})
\mid u_{\vec{q}}(\vec{k}) \rangle.
\end{equation}
It is easy to see that $u_{\vec{q}}(\vec{k})$ is the displacement
of the Fermi surface at the point $\vec{k}$ in the direction of
$\vec{v}_{\vec{k}}$.
Indeed~\cite{negele}, suppose we change the shape of the Fermi surface at
some point $\vec{k}$ by an amount $u(\vec{k})$. The occupation number
changes up to leading order by $\delta \langle n_{0}(\vec{k})\rangle =
-\frac{\partial \langle n_{0}(\vec{k})\rangle}
{\partial \vec{k}} u(\vec{k}) = v_{\vec{k}} \delta(\mu-\epsilon_{\vec{k}})
u(\vec{k})$ which is precisely the quantity which appears in (4.13).
Hence, the coherent states of (4.10) represent {\it deformed Fermi
surfaces} parametrized by the bosonic field $u_{\vec{q}}(\vec{k})$.

Next we define a many-body state which is a direct product of
the coherent states defined above,
\begin{equation}
\mid \{u\} \rangle = \prod_{\vec{k}} \otimes U(\vec{k}) \mid FS \rangle =
\Xi[u] \mid FS \rangle
\end{equation}
where, due to the commutation relation at different $\vec{k}$'s,
\begin{equation}
\Xi[u] = \exp\left(\sum_{\vec{k},\vec{q}} \frac{v_{\vec{k}}}
{2 \mid \vec{q}.\vec{v}_{\vec{k}} \mid} \, u_{\vec{q}}(\vec{k}) \,
a^{\dag}_{\vec{q}}(\vec{k})\right).
\end{equation}
The adjoint is simply,
\begin{equation}
\Xi^{\dag}[u] = \exp\left(\sum_{\vec{k},\vec{q}} \frac{v_{\vec{k}}}
{2 \mid \vec{q}.\vec{v}_{\vec{k}} \mid} \, u^{\ast}_{\vec{q}}(\vec{k}) \,
a_{\vec{q}}(\vec{k})\right).
\end{equation}

{}From the above equations we obtain the overlap of two of these
coherent states
\begin{equation}
\langle \{w\}\mid \{u\} \rangle = \langle FS \mid \Xi^{\dag}[w]
\Xi[u] \mid FS \rangle
= \exp\left(\sum_{\vec{k},\vec{q}} \frac{v_{\vec{k}}^2 \delta(\mu -
\epsilon_{\vec{k}}) }
{2 \mid \vec{q}.\vec{v}_{\vec{k}}\mid} \, w^{\ast}_{\vec{q}}(\vec{k})
 \, u_{\vec{q}}(\vec{k})\right) .
\end{equation}

It is also possible to find the resolution of the identity for this
Hilbert space,
\begin{eqnarray}
1 = \int...\int \prod_{\vec{k},\vec{q}} \left( \frac{v^2_{\vec{k}} \delta(\mu
-\epsilon_{\vec{k}})}{2 \pi \, \mid \vec{q}.\vec{v}_{\vec{k}}\mid} \,
du_{\vec{q}}(\vec{k}) \, du^{\ast}_{\vec{q}}(\vec{k})
\mid u_{\vec{q}}(\vec{k}) \rangle \langle u_{\vec{q}}(\vec{k}) \mid \right)
\nonumber \\
\exp\left(-\sum_{\vec{k},\vec{q}} \frac{v_{\vec{k}}^2
\delta(\mu-\epsilon_ {\vec{k}})}
{2 \mid \vec{q}.\vec{v}_{\vec{k}}\mid} \, \mid u_{\vec{q}}(\vec{k})\mid^2
\right)
\end{eqnarray}
and we conclude, as expected, that they are overcomplete~\cite{klauder}.

In order to study the dynamics of these modes
we can construct, from (4.17) and (4.18), a generating functional as a sum
over the histories of the Fermi surface in terms of
these coherent states in the form $Z = \int D^2u \, \exp\{i \, S[u]\}$ where
$S$ is the action whose lagrangian density is given by ($\hbar=1$),
\begin{equation}
L[u] = \sum_{\vec{k},\vec{q}} \frac{v_{\vec{k}}^2
\delta(\mu-\epsilon_{\vec{k}})
}{2 \, \mid \vec{q}.\vec{v}_{\vec{k}}\mid} \, i \,
u^{\ast}_{\vec{q}}(\vec{k},t)
\frac{\partial u_{\vec{q}}(\vec{k},t)}{\partial t} -
\frac{\langle \{u\}\mid H \mid \{u\} \rangle}{\langle \{u\}\mid \{u\} \rangle}
\end{equation}
where $H$ is the hamiltonian written in the restricted Hilbert space
which will be studied in the next section.

\section{The bosonic hamiltonian}

Consider a fermionic system described by the an hamiltonian of the
form $H = K + U$ where,
\begin{equation}
K=\sum_{\vec{p}} \epsilon_{\vec{p}} \, \, c^{\dag}_{\vec{p}} c_{\vec{p}},
\end{equation}
is the non-interacting term and
\begin{equation}
U =\frac{1}{2 V}
\sum_{\vec{p},\vec{p'},\vec{q}} \, f_{\vec{p}-\vec{q},\vec{p'}+\vec{q}}
\, \, \, c^{\dag}_{\vec{p}+\frac{\vec{q}}{2}} c_{\vec{p}-\frac{\vec{q}}{2}}
c^{\dag}_{\vec{p'}-\frac{\vec{q}}{2}} c_{\vec{p'}+\frac{\vec{q}}{2}}
\end{equation}
is the interaction between the fermions. We will assume that
the interaction between the fermions is important only close to
the Fermi surface. This means that the vectors $\vec{p}$ and
$\vec{p'}$ are at the Fermi surface in what follows.

We  see that the interaction is given in terms of the bosonic
operators as,
\begin{equation}
U=\frac{1}{2 V} \sum_{\vec{p},\vec{p'},\vec{q}}
f_{\vec{p}-\vec{q},\vec{p'}+\vec{q}} \, \, \, n_{-\vec{q}}(\vec{p})
n_{\vec{q}}(\vec{p'}).
\end{equation}

Of course the free part, $K$, can not be written directly in terms
of the bosons since it is quadratic from the beginning. However,
since we are in a restricted Hilbert space what really matters is the
effective dynamics that it can generate. We know that the
operators $n_{\vec{q}}(\vec{k})$
generate the restricted space therefore it is natural to look at
commutation relation
between $K$ and $n_{\vec{q}}(\vec{k})$. Up to first order in $q$ we
find
\begin{equation}
\left[K,n_{\vec{q}}(\vec{k})\right] = -\vec{q}.\vec{v}_{\vec{k}}
n_{\vec{q}}(\vec{k}),
\end{equation}
observe that in order to keep the vector $\vec{k}$ at the Fermi
surface we have to multiply the above expression by a distribution
function of the form $\delta (\mu-\epsilon_{\vec{k}})/ N(0) V$ where
\begin{equation}
N(0) = \frac{1}{V} \sum_{\vec{k}} \delta (\mu-\epsilon_{\vec{k}})
\end{equation}
is the density of states at the Fermi surface.

Now we notice that there is another operator which gives the same commutation
as in (5.4) in the restricted Hilbert space, namely, using (4.5) we find,
\begin{equation}
\left[\sum_{\vec{p},\vec{q'}} n_{-\vec{q'}}(\vec{p}) n_{\vec{q'}}(\vec{p}),
n_{\vec{q}}(\vec{k})\right] = - 2 \, \vec{q}.\vec{v}_{\vec{k}} \,
n_{\vec{q}}(\vec{k}) \, \delta (\mu-\epsilon_{\vec{k}}),
\end{equation}
which means that in the restricted Hilbert space we can rewrite,
\begin{equation}
K = \frac{1}{2 N(0) V} \sum_{\vec{k},\vec{q}} n_{-\vec{q}}(\vec{k})
n_{\vec{q}}(\vec{k}).
\end{equation}

Using the definition (4.6) and the equations (5.3) and (5.7) we conclude
that, in the restricted
Hilbert space, the hamiltonian for the interacting fermions in simply given by,
\begin{equation}
H= \frac{1}{2 \,V} \sum_{\vec{k},\vec{k'},\vec{q},sgn(q)>0}
G^{\vec{q}}_{\vec{k},\vec{k'}}
a^{\dag}_{\vec{q}}(\vec{k}) \, a_{\vec{q}}(\vec{k'}),
\end{equation}
where
\begin{equation}
G^{\vec{q}}_{\vec{k},\vec{k'}} = \frac{\delta_{\vec{k},\vec{k'}}}{N(0)}
\, + f_{\vec{k}-\vec{q},\vec{k'}+\vec{q}},
\end{equation}
which describes a free, quadratic, theory.

It is worth noticing the similarity between (5.8) and (5.9) with
Pomeranchuk's
expression (2.8). This resemblance is expect since we are looking
at the same kind of fluctuations. We would say
that we have a quantum version of Pomeranchuk's construction.
Moreover, all the scaling arguments of the section II can be
used here in other to understand the stability of the Fermi
liquid as a renormalization group fixed point.

\section{Fermi liquid properties}

This section is devoted to show that we can get {\it all} the results of
the Fermi liquid theory from the point of view of the bosons.
In particular, the relationship between the bare mass of the
fermions and the effective mass of the Landau theory is
discussed in appendix C.
We start by studying the thermodynamics (imaginary time) of the system
and then we study its dynamics (real time) from the point
of view of our path integral.

Firstly we will rewrite the lagrangian in (4.19) using the hamiltonian
we have just derived in the last section. Namely, we substitute (5.8)
in (4.19) and use the eigenvalue equation (4.13) in order to get,
\begin{equation}
L[u] = \sum_{\vec{k},\vec{k'},\vec{q},sgn(q)>0} \frac{v_{\vec{k}}^2
\delta(\mu-\epsilon_{\vec{k}})}{2 \, \vec{q}.\vec{v}_{\vec{k}}}
\, u^{\ast}_{\vec{q}}(\vec{k},t) \left(i \delta_{\vec{k},\vec{k'}}
\frac{\partial}{\partial t} - \frac{1}{V} \, W^{\vec{q}}_{\vec{k},\vec{k'}}
\delta(\mu-\epsilon_{\vec{k'}})\right) u_{\vec{q}}(\vec{k'},t)
\end{equation}
where
\begin{equation}
W^{\vec{q}}_{\vec{k},\vec{k'}}= \frac{\vec{q}.\vec{v}_{\vec{k}}
\, \, v_{\vec{k'}}}{v_{\vec{k}}} G^{\vec{q}}_{\vec{k},\vec{k'}}.
\end{equation}

Observe that the vectors $\vec{k}$ and $\vec{k'}$ in the above expressions
are restricted to the Fermi surface due to the Dirac's delta function.
We therefore parametrize these vectors by the solid angles $\Omega$ and
$\Omega'$, respectively. We will assume throughout this paper that the
density of states is a smooth function
across the Fermi surface (which means that there is no Van Hove singularity
present). In our language we make the following substitution,
\begin{equation}
\sum_{\vec{k}} \delta (\mu-\epsilon_{\vec{k}}) f_k =
\frac{N(0) V}{S_d} \int d\Omega f(\Omega)
\end{equation}
where $S_d$ is the integral over the whole solid angle, $S_d = \int d\Omega$.
Furthermore, using the fact that the density of states is smooth we can
replace the velocity by
\begin{equation}
\vec{v}_{\vec{k}} = v_F \vec{n}_{\vec{k}}
\end{equation}
where $\vec{n}_{\vec{k}}$ is an unit vector perpendicular to the Fermi surface
at the point $\vec{k}$.

With these substitutions we can rewrite the lagrangian (6.1) as,
\begin{equation}
L=\frac{N(0) V}{S_d} \int d\Omega \int d\Omega' \sum_{\vec{q}, sgn(q)>0}
\frac{v_F}{2 \vec{q}.\vec{n}(\Omega)}
u^*_{\vec{q}}(\Omega) \left( i \delta_{\Omega,\Omega'} \frac{\partial}
{\partial t} - W_{\vec{q}}(\Omega,\Omega') \right) u_{\vec{q}}(\Omega')
\end{equation}
where
\begin{equation}
W_{\vec{q}}(\Omega,\Omega') = v_F \vec{q}.\vec{n}(\Omega)
\left(\delta_{\Omega,\Omega'}+ \frac{F{\vec{q}}(\Omega,\Omega')}{S_d}\right)
\end{equation}
and $F_{\vec{k},\vec{k'}} = N(0) f_{\vec{k},\vec{k'}}$.

Observe that the field $u_{\vec{q}}(\Omega)$ has dimensions of momentum,
as required by our interpretation, however, we can rewrite the action
in terms of a new dimensionless bosonic field, $\phi_{\vec{q}}(\Omega)$,
by a simple scaling of the original field as,
\begin{equation}
\phi_{\vec{q}}(\Omega) = \left(\frac{v_F N(0) V}{2 S_d
|\vec{q}.\vec{n}(\Omega)|}\right)^{1/2} u_{\vec{q}}(\Omega)
\end{equation}
and rewrite the generating functional as
\begin{equation}
Z=\int...\int \prod_{\vec{q},\Omega,t,sgn(q)>0}
\left(\frac{d\phi_{\vec{q}}(\Omega,t)
d\phi^{*}_{\vec{q}}(\Omega,t)}{2 \pi}\right) e^{i S[\phi,\phi^*]}
\end{equation}
where
\begin{equation}
S=\int dt \left( \int d\Omega \sum_{\vec{q},sgn(q)>0}
i \phi^{*}_{\vec{q}}(\Omega)
\frac{\partial \phi_{\vec{q}}(\Omega)}{\partial t} -
\int d\Omega \int d\Omega' \sum_{\vec{q},sgn(q)>0} \phi^{*}_{\vec{q}}(\Omega)
M^{\vec{q}}(\Omega,\Omega') \phi_{\vec{q}}(\Omega') \right),
\end{equation}
where
\begin{equation}
M^{\vec{q}}(\Omega,\Omega')=\frac{N(0) v_F}{S_d}
\sqrt{|\vec{q}.\vec{n}(\Omega)|
|\vec{q}.\vec{n}(\Omega')|} \left(\frac{S_d \delta(\Omega-\Omega')}{N(0)}
+f(\Omega,\Omega')\right).
\end{equation}

We can split the field $\phi_{\vec{q}}(\Omega)$ into real and
imaginary part as,
\begin{equation}
\phi_{\vec{q}}(\Omega) = \varphi_{\vec{q}}(\Omega)+ i \eta_{\vec{q}}(\Omega)
\end{equation}
and we trace over the imaginary part in order to get,
\begin{equation}
Z=(det M)^{-1/2} \int...\int \prod_{\vec{q},\Omega,t,sgn(q)>0}
\left[ \left(\frac{i}{\pi}
\right)^{1/2} d\varphi_{\vec{q}}(\Omega,t)\right] e^{i S_{eff}[\varphi]}
\end{equation}
where,
\begin{equation}
S_{eff} = \int dt \int d\Omega \int d\Omega'
\sum_{\vec{q},sgn(q)>0} \left( \frac{\partial
\varphi_{\vec{q}}(\Omega)}{\partial t}
\left(M^{\vec{q}}(\Omega,\Omega')\right)^{-1} \frac{\partial
\varphi_{\vec{q}}(\Omega')}{\partial t} -
\varphi_{\vec{q}}(\Omega)
M^{\vec{q}}(\Omega,\Omega') \varphi_{\vec{q}}(\Omega') \right).
\end{equation}

We proceed further and expand the action in its eigenmodes by rewriting,
\begin{equation}
\varphi_{\vec{q}}(\Omega,t) = \sum_{\lambda} \upsilon^{\lambda}_{\vec{q}}
(\Omega) \Phi^{\lambda}_{\vec{q}}(t)
\end{equation}
where
\begin{equation}
\int d\Omega' M^{\vec{q}}(\Omega,\Omega')\upsilon^{\lambda}_{\vec{q}}(\Omega')
=\omega^{\lambda}_{\vec{q}} \upsilon^{\lambda}_{\vec{q}}(\Omega).
\end{equation}
Equation (6.15) can also be rewritten in terms of the original fields
if we use (6.7) and (6.10),
\begin{equation}
v_F \, \vec{q}.\vec{n}(\Omega) \, \left(u^{\lambda}_{\vec{q}}(\Omega)
+\int \frac{d\Omega'}{S_d} F(\Omega,\Omega')
u^{\lambda}_{\vec{q}}(\Omega')\right)
= \omega^{\lambda}_{\vec{q}} u^{\lambda}_{\vec{q}}(\Omega).
\end{equation}

Using the completeness of the states defined in (6.15) we rewrite the
generating functional as,
\begin{equation}
Z=(det M)^{-1/2} \int...\int \prod_{\vec{q},\lambda,t}\left[
\left(\frac{i}{\pi}
\right)^{1/2} d\Phi^{\lambda}_{\vec{q}}(t)\right] e^{i S_{eff}[\Phi]}
\end{equation}
where
\begin{equation}
S_{eff} = \int dt \sum_{\lambda,\vec{q},sgn(q)>0}
\frac{1}{\omega^{\lambda}_{\vec{q}}}
\left(\left(\frac{\partial \Phi^{\lambda}_{\vec{q}}(t)}{\partial t}\right)^2
-\left(\omega^{\lambda}_{\vec{q}}\right)^2
\left(\Phi^{\lambda}_{\vec{q}}(t)\right)^2\right),
\end{equation}
which, as expected, is the action for a free massive bosonic field.

\subsection{Thermodynamic Properties}

The extension of the lagrangian (6.5) to imaginary time ({\it i.~e.~}
finite temperature) is trivial.
We will consider first the case in which only the diagonal term in
eq~(6.6) is present, namely,
\begin{equation}
W_{\vec{q}}(\Omega,\Omega') = E_{\vec{q}}(\Omega) \delta_{\Omega,\Omega'}
\end{equation}
and therefore it is straightforward to show that the partition function
is given by,
\begin{equation}
Z=exp(-\beta F) = \prod_{\Omega,\vec{q},sgn(q)>0} \left[\sinh\left(
\frac{E_{\vec{q}}(\Omega) \beta}{2}\right)\right]^{-1}
\end{equation}
where $\beta = 1/T$ is the inverse of temperature (our units are
such that $k_B = \hbar= 1$).

Using the expression for the free energy, $F$, given in (6.20) we easily
calculate the specific heat,
\begin{equation}
C_V = \frac{\beta^2}{4} \sum_{\Omega,\vec{q},sgn(q)>0}
\frac{E^2_{\vec{q}}(\Omega)}{\sinh^2\left(\frac{E_{\vec{q}}(\Omega)\beta}
{2}\right)}.
\end{equation}

We have to be careful in order to understand the integrals in (6.21).
We built a reference frame which
runs over the surface (a Fresnel frame as in differential geometry).
The Fermi velocity (normal to the Fermi surface at $\vec{p}_F$)
and gradient of the Fermi velocity (which spans the vectors in
the plane tangent to the the Fermi surface at $\vec{p}_F$)
form a {\it local} orthogonal basis for the vectors $\vec{q}$.
We therefore split the integral in (6.21) in a part which is
normal to the surface, $q_N$, and another which is in the
plane tangent to the surface, $q_P$. In the absence of
interactions the dispersion is given by (6.6) as,
\begin{equation}
E(q_N) = v_F \, q_N .
\end{equation}

Firstly we observe that the density of states is given as an
integral over the tangent component. Indeed, from (5.5) and the
definition of the local frame, we have,
\begin{equation}
N(0)= \frac{1}{S_d} \int \frac{dS}{v_F} = \frac{1}{S_d} \int d\Omega
\int dq_{P}(\Omega) \frac{q_{P}^{d-2}(\Omega)}{v_F}.
\end{equation}
We also have assumed that the normal component is not affected
by the curvature (see (6.4)). Thus we can integrate out the
tangent component in (6.21) using (6.23). The final result is,
\begin{equation}
\frac{C_V}{V} = \frac{\beta^2}{4} N(0) v_F \int_{0}^{\infty} d q_{N}
\frac{E^2(q_{N})}{\sinh^2\left(\frac{E(q_{N})\beta}{2}
\right)}.
\end{equation}

Using (6.22) we finally find,
\begin{equation}
\frac{C_V}{V} = \frac{\pi^2}{3} N(0) T
\end{equation}
which is the expected result for a Fermi liquid \cite{baym}. Observe that
the dimensionality plays no role here except in the calculation of
the density of states.

We can clearly see the differences between the bosonic excitations
described in this paper, which are due to coherent superposition of
particle-hole pairs, and the usual free bosonic excitations. First
of all, naively we would expect that the specific heat should
behave as $T^3$ as in usual free bosonic theories since the
bosonic hamiltonian (5.8) is quadratic in the bosons. However,
the the bosonic field in the Fermi liquid {\it lives}
on the Fermi surface, that is, it is topologically constrained.
The Fermi velocity , through the density of states, defines the
metric of the manifold where the fields propagate. In the case
without interactions the bosonic fields
oscillate without coherence (like decoupled harmonic oscillators
with phases distributed at random). These oscillation are responsible
for the contribution of the particle-hole continuum to the specific
heat as shown in (6.25).

We can proceed further and evaluate the effect of the interactions in
this problem. The general approach would be to calculate the
eigenvalues of equation (6.15) and evaluate the bosonic determinant
in the partition function. However, we use an approach which is
inspired by the calculations of the Fermi liquid theory~\cite{pethick}.
Firstly, we expand the interaction term for
small angles by defining the vector $\vec{p}=\vec{k'}-\vec{k}$ such that
\begin{equation}
f_{\Omega,\Omega'} = a + b \left( \frac{\vec{k}}{k}.\frac{\vec{p}}{p}\right)
\end{equation}
where $a$ and $b$ are parameters which depend on the specific form of
the interaction. Due to the geometry of the interaction we can rewrite
(6.26) as~\cite{houghton}
\begin{equation}
f_{\Omega,\Omega'} = a + b \frac{q_{N}^2}{q_{N}^2 + p^2/4}.
\end{equation}

Since we are working with a local expansion we will replace the non-local
term in (6.6) by a local one which is an average over the surface,
\begin{equation}
f_{\Omega,\Omega'} \to \left(\int d\Omega' f_{\Omega,\Omega'} \right)
\delta_{\Omega,\Omega'}
\to b \sum_{\vec{p}} \frac{q_{N}^2}{q_{N}^2 + p^2/4} \delta_{\Omega,\Omega'},
\end{equation}
changing the sum to an integral and taking into account that
this integral must be done on the surface we found that the
change in the energy due to interactions can be written as
\begin{equation}
\Delta E (q_{N}) = 16 v_F b q_{N}^3 \frac{N(0) S_{d-1}}{p_F^{d-1} S_d}
\int_{0}^{\Lambda} dp \frac{p^{d-2}}{p^2+4q_{N}^2}
\end{equation}
where $\Lambda$ is the cut-off.
In three dimensions, we have,
\begin{equation}
E(q_{N}) = v_F q_{N} - \frac{8 b v_F N(0)}{p_F^2} q_{N}^3 ln\left(
\frac{q_{N}}{\Lambda}\right)
\end{equation}
and in two dimensions one finds,
\begin{equation}
E(q_{N}) = v_F q_{N} + \pi b \frac{v_F N(0)}{p_F} q_{N}^2.
\end{equation}

Observe that these expression depend only on the normal
component and we can use equation (6.24) in order to
calculate the corrections for the specific heat due to
interactions. Up to first order in $b$ we have,
\begin{equation}
\frac{\delta C_V}{V} = - \alpha T^3 ln\left(\frac{T}{\Lambda v_F}\right)
\end{equation}
in three dimensions, and,
\begin{equation}
\frac{\delta C_V}{V} = \eta T^2
\end{equation}
in two dimensions, where,
\begin{equation}
\alpha =\frac{16 p_F^2 b}{15 \pi^2 v_F^4}
\end{equation}
and
\begin{equation}
\eta=\frac{3 \zeta(3) p_F b}{\pi v_F^3},
\end{equation}
where $\zeta(n)$ is the Riemman Zeta function of $n$.

The result (6.32) is well known \cite{baym,pethick} (apart from the
prefactor which is different due to the average over the Fermi surface
we have carried out in (6.28)) and the
result (6.33) was obtained recently using the RPA approximation
\cite{coffey}.

We conclude therefore that in our approximation we are counting
the correct number of states at low temperatures. This result
proves the consistency of our method.
\subsection{The semiclassical dynamics}

Notice that the lagrangian (6.5) is quadratic in the fields and, therefore, the
semiclassical approximation is exact. The semiclassical equations of
motion for these lagrangian
are given by the saddle point equation (the Euler-Lagrange equations)
derived from $L$, namely,
\begin{equation}
i \frac{\partial u_{\vec{q}}(\Omega,t)}{\partial t} =
q v_F \, \cos \theta \, u_{\vec{q}}(\Omega,t) + q v_F \, \cos \theta
\int \frac{d\Omega'}{S_d} \, F(\Omega,\Omega') \, u_{\vec{q}}(\Omega',t)
\end{equation}
where we have defined the angle $\theta$ by $\vec{q}.\vec{n}(\Omega) =
\cos \theta$. Observe that if,
\begin{equation}
u_{\vec{q}}(\Omega,t) = e^{-i \omega^{\lambda}_{\vec{q}} t}
u_{\vec{q}}^{\lambda}(\Omega),
\end{equation}
we recover equation (6.16).

Eq.~(6.36) is the Landau equation of motion
for sound waves (the collective modes) of a neutral Fermi liquid
where $f_{\vec{k},\vec{k'}}$ is the scattering amplitude for
particle-hole pairs ~\cite{baym}. Observe that the Landau equation
gives the eigenmodes of the problem (see (6.15)).

We conclude therefore that our bosons are the sound waves which
propagate around the Fermi surface at zero temperature. The solution of
(6.16) (or (6.36)) will give the
possible values for the frequencies of oscillation for these modes and
they will depend essentially on the Landau parameters of the theory.
The Landau equation, (6.16), yields solutions which represents both
stable collective modes (``sounds") as well as solutions with imaginary
frequency which represent the particle-hole continuum. This
behavior is a direct consequence of the phase space. Notice that in one
dimension these unstable solutions are absent and only the collective
mode is left.

In order to illustrate the behavior of these sound modes we present here
an explicit calculation in two dimensions. This calculation will
enable us to understand many interesting features of these sound modes.

As in the section II, we expand the interaction and the displacement
in Fourier components,
\begin{equation}
F(\theta,\theta') = \sum_{n=-\infty}^{\infty} F_{n} e^{i(\theta-\theta')n}
\end{equation}
where $F_{-n} = F^{*}_{n}=F_{n}$ due to the symmetry of the problem and,
\begin{equation}
u(\theta) = \sum_{n=-\infty}^{\infty} \, u_n \, e^{i \theta n} ,
\end{equation}
where $u_{-n}=u_{n}^{*}$.

{}From (6.16) we have,
\begin{equation}
(s-\cos\theta)u(\theta) = \cos\theta \int_{0}^{2\pi} \frac{d\theta'}{2\pi}
F(\theta,\theta') u(\theta')
\end{equation}
where $v_F s = \omega/q$ is the velocity of the sound waves.

Using (6.38), (6.39) and the orthogonality relation of the Fourier
components it is easy to show that (6.40) can be written as a matrix
equation,
\begin{equation}
(1+F_n) u_{n} = \sum_{m=-\infty}^{\infty} F_{m} K(m-n) u_{m}
\end{equation}
where
\begin{equation}
K(n) = \int_{0}^{2\pi} \frac{d\theta}{2\pi} \frac{s}{s-\cos\theta}
e^{i n \theta}.
\end{equation}
We allow $s$ to have an infinitesimal imaginary part and we integrate
(6.42) in the complex plane in a contour in the unit circle
around the origin. We found
\begin{eqnarray}
K(n) = \frac{|s|}{\sqrt{s^2-1}} (sgn(s))^n \left(s- \sqrt{s^2-1}
\right)^{|n|}
\hspace{1cm} \, |s|>1
\nonumber
\\
=\frac{s}{i \sqrt{1-s^2}} \left(s- i \sqrt{1-s^2}
\right)^{|n|}
\hspace{1cm} \, |s|<1.
\end{eqnarray}

It is easy to see from (6.43) that all the modes are damped if the
velocity is smaller of the Fermi velocity, $|s|<1$. This result is
clearly expected, it is nothing but the Landau damping of the
collective modes~\cite{baym}.

Here we are only interested in the stable solutions of (6.41).
Therefore, for $s>1$ we can replace (6.41) by
\begin{equation}
(1+F_n)u_n = \sum_{m=-\infty}^{\infty} F_m K(m-n) u_m
\end{equation}
where we define a kernel,
\begin{equation}
K(n) = A(s) e^{-|n| \alpha(s)}
\end{equation}
with,
\begin{equation}
A(s) = \frac{s}{\sqrt{s^2-1}}
\end{equation}
and
\begin{equation}
\alpha(s) = ln\left(s + \sqrt{s^2-1}\right).
\end{equation}

For a purely local interaction in the real space the interaction is
angle independent (long range over the surface in momentum space)
and we define,
\begin{equation}
F_n = F_0 \delta_{n,0}.
\end{equation}
Substituting (6.48) in (6.44) for the $n=0$ we obtain the allowed value
of $s$,
\begin{equation}
s=\frac{1+\frac{1}{F_0}}{\sqrt{\left(1+\frac{1}{F_0}\right)^2 -1}}
\end{equation}
and for $n \neq 0$ we obtain the value of the Fourier components
\begin{equation}
u_n = F_0 K(n) u_0.
\end{equation}
Substituting (6.50) in (6.39) and summing up an ordinary geometric series
with a little help of (6.49) we easily get,
\begin{equation}
u(\theta) = u_0 F_0 \, \, \frac{\cos\theta}{s-\cos\theta}
\end{equation}
which is expect result if we have just solved (6.40) for a
constant interaction. The result (6.51) is the well known
zero sound mode which has the same shape as the zero sound in
the three dimensional case~\cite{baym}.

We can explore even more the matrix equation (6.44) due to the
simplicity of the kernel in (6.45). We are mainly interested in the
behavior of the Fermi surface in the forward scattering direction
since it has been argued \cite{anderson} that it is in this direction
that pathological effects can happen.

Since we are not interested in the behavior over the whole Fermi
surface we will transform the matrix equation (6.44) in a
simple second order differential equation by expanding the kernel
in terms of the Fermi wavelength.

Consider the arc-length defined on the Fermi surface by,
\begin{equation}
\nu = p_F \theta,
\end{equation}
the expansion (6.39) is rewritten as,
\begin{equation}
u(\nu) = \sum_{\lambda} e^{i \lambda \nu} u(\lambda)
\end{equation}
and due to the periodic boundary conditions,
$u(\nu + 2\pi p_F) = u(\nu)$, we must have,
\begin{equation}
\lambda_n = \frac{n}{p_F}.
\end{equation}

Observe from (6.54) that in the limit of $p_F \to \infty$
the matrix equation (6.44) can be replace by an integral equation,
\begin{equation}
(1+F(\lambda)) u(\lambda) = \int_{-\infty}^{+\infty} d\lambda'
G(\lambda-\lambda') F(\lambda') u(\lambda')
\end{equation}
where
\begin{equation}
G(\lambda) = 2 \frac{A}{\alpha} \int_{-\infty}^{+\infty} \frac{dx}{2\pi}
\frac{e^{ i x \lambda}}{1+\left(\frac{x}{p_F \alpha}\right)^2}
\end{equation}
where we have used (6.45) and an integral representation for the
exponential function. Expanding (6.56) up to second order in $p_F^{-1}$
one finds,
\begin{equation}
G(\lambda) = 2 \frac{A}{\alpha} \left( \delta(\lambda)
+\frac{1}{\left(p_F \alpha\right)^2}
\frac{d^2 \delta(\lambda)}{d\lambda^2}\right).
\end{equation}
If we substitute (6.57) in the original equation (6.55) we find a
differential equation which should be valid for small arc-lengths
compared to $2 \pi p_F$. We proceed further and define the following
quantity,
\begin{equation}
z(\lambda) = F(\lambda) u(\lambda)
\end{equation}
which, from the arguments given above, obeys the differential equation,
\begin{equation}
\left(-\frac{d^2 }{d\lambda^2} + U(s,\lambda)\right) z(\lambda)=
\epsilon(s) z(\lambda)
\end{equation}
where
\begin{equation}
U(s,\lambda)=\frac{\left(p_F \alpha(s)\right)^2}{F(\lambda)}
\end{equation}
and
\begin{equation}
\epsilon(s)= \left(p_F \alpha(s)\right)^2 \left(2 \frac{A(s)}{\alpha(s)}
-1\right).
\end{equation}

Eq.~(6.59) has the form of a time independent Schr\"odinger equation
for a particle in a potential $U$ with energy $\epsilon$.
Once we know the boundary conditions for the problem we can
determine its eigenfunctions (and therefore we determine $u$ using
(6.58)) and its eigenenergies (and we determine $s$ using (6.61)).

In this paper we will assume a particularly simple form for the
interaction term which has been used in the literature in order to
investigate the problem of interactions of finite range~\cite{nozieres2}
\begin{equation}
F(\nu) = \pi \gamma p_F \lambda_c e^{-\lambda_c |\nu|}.
\end{equation}
There are two parameters here, $\gamma$ gives the strength of
the interaction and $\lambda_c$ plays the role of the range of
the interaction. When $\lambda_c \to 0$ the interaction is angle independent
(is short range in the real space) and we obtain the zero sound
solution already discussed; when $\lambda_c \to \infty$ the interaction
is long range in real space and strongly angle dependent in momentum
space. This parameter will enable us to study the crossover from
short range to long range interaction in terms of the dynamics of the
sound waves.

The Fourier transform of (6.62) is given by,
\begin{equation}
F(\lambda) = \gamma \frac{\lambda^2_c}{\lambda^2+\lambda_c^2},
\end{equation}
which is a lorentzian.

Using (6.63) in (6.59) we obtain the equation for the one-dimensional
harmonic oscillator,
\begin{equation}
\left( -\frac{d^2}{d\lambda^2} + \beta^4(s) \lambda^2\right) z(\lambda)
= E(s) z(\lambda)
\end{equation}
where
\begin{equation}
\beta(s) = \left(\frac{p_F \alpha(s)}{\gamma^{1/2} \lambda_c}\right)^{1/2}
\end{equation}
and
\begin{equation}
E(s) = \epsilon(s) - \left(\frac{p_F \alpha(s)}{\gamma^{1/2}}\right)^2.
\end{equation}
Of course the natural boundary condition is that the function $z$ vanishes
at the infinity.

The solution of (6.64) is standard. The eigenfunctions are,
\begin{equation}
z_{n}(\lambda) = e^{-\beta^4 \lambda^2} H_{n}(\beta^2 \lambda)
\end{equation}
where $H_n$ is the Hermite polynomial of order $n$ and the
eigenenergies are,
\begin{equation}
E(s_n) = 2 \beta^2(s_n) \left(n+\frac{1}{2}\right).
\end{equation}

Equation (6.68) is self-consistent and it gives the allowed values of $s$.
If we substitute (6.46), (6.47) in (6.68) we get the following transcendental
equation,
\begin{equation}
(1+\frac{1}{\gamma}) \alpha(s_n) +
2 \left[\frac{1}{\gamma^{1/2} p_F \lambda_c}\left(n+\frac{1}{2}\right)
-A(s_n)\right]=0.
\end{equation}
It is very easy to check that the value of $s_n$ decreases monotonically
with $n$ and, in particular, $lim_{n\to\infty}s_n=1$. This result
means that the modes approach asymptotically the particle-hole
continuum.

Furthermore, substituting (6.67) and (6.63) in (6.58) we find
\begin{equation}
u_n(\lambda)= (\lambda^2 + \lambda^2_c) e^{-\beta^4 \lambda^2}
H_{n}(\beta^2 \lambda).
\end{equation}
We now Fourier transform (6.70) back to the arc-length in order to get,
\begin{equation}
u_n(\nu)= \left( 2n+1+\beta^2(s_n) \lambda_c -\left(\frac{\nu}{\beta(s_n)}
\right)^2\right) e^{-\frac{1}{2} \left(\frac{\nu}{\beta(s_n)}\right)^2}
H_{n}\left(\frac{\nu}{\beta(s_n)}\right).
\end{equation}

Equations (6.69) and (6.71) are the solutions of (6.44) in the limit
when the Fermi momentum is much larger than the momentum transfer
between the particle-hole pairs (which means, as we said, that we
are looking close to the forward scattering direction).

We want to change the range of the interaction, $\lambda_c$, but we
keep the density of the system constant. In this
case the product $p_F \lambda_c$ is finite ( $\lambda_c p_F \sim
\frac{a_0}{r_0}$ where $r_0$ is the mean distance
between the particles and $a_0$ is the Bohr radius). In other words,
it means that $\beta \lambda_c$ is finite in
what follows. Therefore, keeping the above product
constant and sending $\lambda_c \to 0$ we find from (6.65),
\begin{equation}
\beta \to \infty
\end{equation}
and from (6.71),
\begin{equation}
u_n(\nu) \to constant,
\end{equation}
if $n$ is even and $u_n(\nu) = 0$ if $n$ is odd.
Observe that this result is in agreement with the zero sound
result, (6.51), which is constant in the forward direction ($\theta
\sim 0$).

However, for $\lambda_c \to \infty$ it is easy to see that,
\begin{equation}
\beta \to 0,
\end{equation}
and
\begin{equation}
u_n(\nu) \sim \left(\frac{\nu}{\beta}\right)^2
e^{-\frac{1}{2} \left(\frac{\nu}{\beta}\right)^2}
H_{n}\left(\frac{\nu}{\beta}\right)
\end{equation}
which oscillates strongly near $\nu =0$.
We would argue therefore that, while for short range interactions
($\lambda_c \to 0$) the field $u(\nu)$ is smooth around the
forward scattering direction, for long range interactions
($\lambda_c \to \infty$) the field becomes rough near the same
direction. This could represent a signal that there is something
particularly non-conventional happening there. Notice, on one hand,
that this result has similarities with the problem of long-range order
in low dimensional systems~\cite{mermim} since it is well known that
one-dimensional systems
at zero temperature with local interactions do not have long range order.
However, on the order hand, also quantum one-dimensional interfaces at
zero temperature
can not have long range order while two-dimensional interfaces always have
long range order~\cite{fradkin}. Since the Fermi surface has the properties
of a real interface in momentum space it could be argued that the Fermi
surface could undergo a roughening transition of the Kosterlitz-Thouless
type~\cite{kosterlitz} for singular interactions.

However, the calculations we have carried out are purely classical
and up to now we do not know if quantum fluctuations can wash out
this behavior. It is worth mention that in quantum crystals the quantum
fluctuations can wash out the classical ones~\cite{fradkin}.

\subsection{Quantum fluctuations of the Fermi surface}

It was shown some time ago~\cite{mermim} that quantum one-dimensional
crystals can lose long range order when the interactions are
sufficiently local. Quantitatively
it means that the equal-time correlation function for the deviations of the
equilibrium position of the atoms, $u(R)$, diverges logarithmically
for long distances, that is, $\langle u(R)u(R') \rangle \sim
ln|R-R'|$ as $|R-R'| \to \infty$. We would argue, due to the classical
calculations we have carried out that the same could
happen for the fields $u_{\vec{q}}(\Omega)$ for long range interactions
in real space. Therefore, in order to understand this problem we calculate
the quantum correlation function for different pieces of the Fermi surface.

{}From (6.18) is straightforward to calculate the correlation function for
different pieces of the Fermi surface, namely,
\begin{equation}
\langle T \varphi_{\vec{q}}(\Omega,t) \varphi_{\vec{q'}}(\Omega',t')
\rangle = \sum_{\lambda,\lambda'} \upsilon^{\lambda}_{\vec{q}}(\Omega)
\upsilon^{\lambda'}_{\vec{q'}}(\Omega') \, \, \langle T
\Phi^{\lambda}_{\vec{q}}(t)
\Phi^{\lambda'}_{\vec{q'}}(t')\rangle
\end{equation}
where $T$ is the chronological operator and
$\upsilon^{\lambda}_{\vec{q}}(\Omega)$
are the solutions of (6.15). The correlation function in the r.h.s. of
(6.76) is easily calculated from (6.18) (we assume that
the eigenfrequencies $\omega_{\vec{q}}^{\lambda}$ are well behaved functions
of $\vec{q}$ and ${\lambda}$) and it reads,
\begin{equation}
\langle T \Phi^{\lambda}_{\vec{q}}(t)
\Phi^{\lambda'}_{\vec{q'}}(t')\rangle = \delta_{\lambda,\lambda'}
\delta_{\vec{q},\vec{q'}} e^{ i \omega^{\lambda}_{\vec{q}} |t-t'|}.
\end{equation}

Thus from (6.76) and (6.77),
\begin{equation}
\langle T \varphi_{\vec{q}}(\Omega,t) \varphi_{\vec{q'}}(\Omega',t')
\rangle = \delta_{\vec{q},\vec{q'}} \, \,
\sum_{\lambda} \upsilon^{\lambda}_{\vec{q}}(\Omega)
\upsilon^{\lambda}_{\vec{q}}(\Omega') \,  \,
e^{i \omega^{\lambda}_{\vec{q}} |t-t'|}
\end{equation}
and therefore the equal-time correlation function is simply
(use the completeness of the states in (6.15)),
\begin{equation}
\langle T \varphi_{\vec{q}}(\Omega,t) \varphi_{\vec{q'}}(\Omega',t)
\rangle = \delta_{\vec{q},\vec{q'}} \delta(\Omega-\Omega').
\end{equation}

Eq. (6.79) shows that the correlation function is always finite
and therefore there is no roughening (there is
long range order), that is, the Fermi surface is always smooth.

The comparison between this result and the result of quantum
crystals~\cite{fradkin} we know that this result is due to a
conservation law, namely, the local conservation of the volume of
the Fermi surface (the Luttinger's theorem~\cite{luttinger}).
{}From the other point of view this is expect since the sound
waves obey a diffusion like equation, (6.36), and therefore
a bump on the Fermi surface diffuses leaving nothing behind it.

\section{Conclusions}

We construct a bosonization of a Fermi liquid in any number of
dimensions in the limit of long wavelengths.
The bosonization is valid for a restricted set of states
close to the Fermi surface. We generate a set of creation and
annihilation operators which span all the restricted Hilbert
space. We showed that is possible to
construct a set of coherent states (bosonic shape fields)
which are coherent
superposition of particle-hole excitation close to the Fermi
surface and which are the displacements of the
Fermi surface in some direction. The physical interpretation
for the existence of these fields comes from the observation
that the Fermi surface is a real quantum object which
is responsible for the whole physics of Fermi liquid
systems at low energies.
{}From the coherent states we generate all the thermodynamics
and semi-classical dynamics of a Fermi liquid.

We have shown that an interacting
hamiltonian of the original fermions is simply quadratic in
terms of the bosons, which means that the bosons are free and
move on the Fermi surface like sound waves. They are not
free fields in the usual sense since they are topologically
constrained to live in the Fermi surface. This new feature
produces very interesting results. And, in particular,
reproduces {\it all} the results of the Fermi liquid
theory from a bosonic point of view.

We obtain the correct thermodynamics of a Fermi liquid
even with the
corrections due to the scattering between the fermions.
This observation lead us to conclude that our approach
is consistent and that we are counting the correct number of
states in the Hilbert space.

Moreover, we have shown that
the semiclassical dynamics of these sound waves is described
by the Landau theory. In two dimensions we calculate the
form of these sound waves explicitly for short range interactions
(zero sound) and long range interactions. We have shown that, at
classical level, we could conclude that it is possible to have
a roughening like transition of the Kosterlitz-Thouless type, which
is closely related with the absence of long range order in one
dimensional crystals at zero temperature. However, we show that
quantum fluctuations destroy this effect due to the local conservation
of the volume of the Fermi surface (Luttinger's theorem).

\acknowledgements

We gratefully thank Gordon Baym for many illuminating conversations,
for his encouragement and sharp criticism. We benefitted from an
early conversation with F.~D.~M.~Haldane and we wish to thank him for
explaining to one of us (E.~F.~) his (still unpublished) picture of the
Fermi surface as a quantum mechanical extended object. We also
acknowledge Brad Marston for interesting discussions during the
APS March Meeting in Seattle (1993).
A.~H.~C.~N. thanks Conselho Nacional de Desenvolvimento Cient\'ifico
e Tecnol\'ogico, CNPq, (Brazil) for a scholarship and J.~M~.P~.Carmelo
for useful discussions.
This work was supported in part by NSF Grant DMR91-22385 at the
University of Illinois at Urbana-Champaign.

\newpage

\appendix

\section{Fermi liquid and non-Fermi liquid behavior}

In this appendix we discuss the possibility of other behaviors
for the imaginary part of the self energy than the conventional
Fermi liquid one, (3.13). We will assume yet that the real part
of the self energy is linear in the frequency close to the Fermi
surface but for the imaginary part we will assume that,
\begin{equation}
\Sigma_I(\vec{k},w) = C_{\vec{k}} \, |w - \mu|^{1+\nu} \,
sgn(\mu - w).
\end{equation}

If we substitute (A1) in (3.12) and make the same change of
variables which lead to (3.14) we find,
\begin{equation}
E^{s}_{\vec{k}}=\int^{\infty}_{\frac{\mid\delta k\mid \mid \nabla
\epsilon_{\vec{p}_F}\mid}{\zeta}} \frac{dx}{2 \pi} \,
\left(\mu+\epsilon^{0}_{\vec{k}}-\frac{1}{x} \mid\delta k\mid
\mid \nabla \epsilon_{\vec{p}_F}\mid\right)
\frac{C_{\vec{k}} x^{\nu-1} \mid\delta k\mid^{\nu}
\mid \nabla \epsilon_{\vec{p}_F}\mid^{\nu}}
{\left(Z_{\vec{k}}^{-2} x^{2\nu} (x+sgn(\delta k))^2 + C_{\vec{k}}^2
\delta k^{2\nu} \mid \nabla \epsilon_{\vec{p}_F}\mid^{2\nu} \right)}.
\end{equation}

When $\nu>0$ it is easy to calculate the limit analogous to (3.15),
\begin{equation}
Lim_{\delta k \to 0} \, \frac{C_{\vec{k}} \mid\delta k\mid^{\nu}
\mid \nabla\epsilon_{\vec{p}_F}\mid^{\nu}}
{\left(Z_{\vec{k}}^{-2} x^{2\nu} (x+sgn(\delta k))^2 +
C_{\vec{k}}^2 \delta k^{2 \nu}
\mid \nabla \epsilon_{\vec{p}_F}\mid^{2 \nu}\right)} =
\pi \mid Z_{\vec{p}_F} \mid
\delta\left(x^{\nu}(x+sgn(\delta k)\right),
\end{equation}
which gives the same result as in (3.16) (Fermi liquid behavior).
However, if $\nu<0$ the above limit takes a different form, namely,
\begin{equation}
Lim_{\delta k \to 0} \, \frac{C_{\vec{k}} \mid\delta k\mid^{\nu}
\mid \nabla\epsilon_{\vec{p}_F}\mid^{\nu}}
{\left(Z_{\vec{k}}^{-2} x^{2\nu} (x+sgn(\delta k))^2 +
C_{\vec{k}}^2 \delta k^{2 \nu}
\mid \nabla \epsilon_{\vec{p}_F}\mid^2\right)} =
\frac{1}{C_{\vec{k}} |\delta k|^{\nu} \, |\nabla
\epsilon^0_{\vec{p}_F}|^{\nu}},
\end{equation}
and therefore, substituting in (A2) we find,
\begin{equation}
E^{s}_{\vec{k}} = \left(\frac{1}{2(1-\nu)} C_{\vec{k}}^{\nu-2}
|\delta k |\nabla \epsilon^0_{\vec{p}_F}||^{(\nu-1)^2}\right)
\Theta(-\delta k).
\end{equation}
The discontinuity is easily calculated in the limit as $\delta k \to 0$,
\begin{equation}
\Delta E(\vec{p}_F) = \left(\frac{C_{\vec{p}_F}^{\nu-2}
|\nabla \epsilon^0_{\vec{p}_F}|^{(\nu-1)^2}}{2(1-\nu)}\right)
|\delta k|^{(\nu-1)^2} \to 0,
\end{equation}
and therefore there is no singularity left (non-Fermi liquid behavior).

\section{Bosonization in one spatial dimension}

In this appendix we show that the formulae obtained in section IV
of our article are consistent with the bosonization procedure in
one dimension. In order to do so we prove that we can obtain the
same commutation relations for the densities
as it is obtained in the standard procedures of bosonization.
Therefore, all other results relative to the calculation of the
fermionic operators and the one particle Green's function follows
from this result.

We will concentrate particularly on the Tomonaga's model \cite{tomonaga}
but the extension of the arguments to the Luttinger's model \cite{authors}
is absolutely straightforward.

The Tomonaga's model for spinless electrons is described by the following
hamiltonian,
\begin{equation}
H = \sum_{p} \epsilon_{p} c^{\dag}_{p} c_{p} +
\frac{1}{2L} \sum_{q} V_{q} \rho(q) \rho(-q)
\end{equation}
where the dispersion relation is given by,
\begin{equation}
\epsilon_{p} = v_F \mid p \mid \, ,
\end{equation}
$V_{q}$ is Fourier transform of the electron-electron interaction,
$L$ is the length of the system and
\begin{equation}
\rho(q) = \sum_{k} n_{q}(k)
\end{equation}
is the density operator.

Now we split the density operator in two terms, one for right
movers and other for left movers,
\begin{eqnarray}
\rho_1(q) = \sum_{k>0} n_{q}(k)
\nonumber
\\
\rho_2(q) = \sum_{k<0} n_{q}(k),
\end{eqnarray}
thus, $\rho(q) = \rho_1(q)+\rho_2(q)$.

The one dimensional version of the commutation relation (4.5) is,
\begin{equation}
\left[n_{q}(k),n_{-q'}(k')\right] = sgn(k) \,
\delta_{k,k'} \delta_{q,q'} \, q v_F \, \delta(\mu- v_F \mid k \mid).
\end{equation}
Therefore we have from (B4) we easily find,
\begin{equation}
\left[\rho_i(q),\rho_j(-q')\right] = (-1)^{j+1} \delta_{i,j} \, q v_f
\, \sum_{k(-1)^j<0} \delta(\mu-v_F \mid k \mid).
\end{equation}
where the chemical potential is written as $\mu = v_F p_F$.

In the thermodynamic limit we transform the sum into an
integral ($\sum_k \to \frac{L}{2 \pi} \int dk$) and we finally get,
\begin{equation}
\left[\rho_i(q),\rho_j(-q')\right] = \delta_{i,j} (-1)^{j+1}
\frac{q L}{2 \pi} \, ,
\hspace{1cm} i,j =1,2 \, \, ,
\end{equation}
as expected for spinless fermions \cite{tomonaga}.

The bosonic operators are defined in terms of the densities
for $q>0$ as (compare with (6.7))
\begin{eqnarray}
b_{q} = \sqrt{\frac{2 \pi}{q L}} \rho_1(q)
\nonumber
\\
b_{-q} = \sqrt{\frac{2 \pi}{q L}} \rho_2(-q)
\end{eqnarray}
with their adjoint they obey the usual bosonic algebra,
\begin{equation}
\left[b_k,b^{\dag}_{k'}\right] = \delta_{k,k'}.
\end{equation}

Following the same kind of argument of section V we can prove
that the Tomonaga's model is purely quadratic in terms of the
bosons and can be easily diagonalized by a Bogoliubov transformation.
In particular the calculation of the correlation functions follows
exactly as in the work of Mattis and Lieb \cite{authors} and Luther
and Peschel \cite{luther2} for the Luttinger's model.

\section{The effective mass}

In this appendix  we show the complete consistency of our approach and
the Landau theory of the Fermi liquid via the calculation of the
effective mass of the Landau theory in terms of the bare electronic
mass.

Suppose we displace all pieces of the Fermi surface in momentum space
by the same infinitesimal amount, $\vec{q}$. This is equivalent
to look to the system from a referential which moves relative to it
with constant velocity. By simple geometric arguments it is easy to
see that the displacements of the Fermi surface will change by the quantity,
\begin{equation}
u_{\vec{q}}(\vec{k}) \to u_{\vec{q}}(\vec{k}) +
\frac{\vec{q}.\vec{v}_{\vec{k}}}
{v_{\vec{k}}}.
\end{equation}

The total change in the action can be obtained from (6.1).
Up to first order in $q$ one gets,
\begin{equation}
\delta L = -\sum_{\vec{k},\vec{k'},\vec{q}} \frac{v^2_{\vec{k}} \delta(\mu-
\epsilon_{\vec{k}})}{V \vec{q}.\vec{v}_{\vec{k}}} u^{*}_{\vec{q}}(\vec{k})
\left(W^{\vec{q}}_{\vec{k},\vec{k'}} \delta(\mu-\epsilon_{\vec{k'}})
\frac{\vec{q}.\vec{v}_{\vec{k'}}}{v_{\vec{k'}}}\right).
\end{equation}
Using (6.2) and (5.9) we can rewrite the above expression as,
\begin{equation}
\delta L = -\sum_{\vec{k},\vec{q}} v_{\vec{k}} \delta(\mu-\epsilon_{\vec{k}})
u^{*}_{\vec{q}}(\vec{k}) \left(\vec{q}.\vec{J}_{\vec{k}}\right),
\end{equation}
where,
\begin{equation}
\vec{J}_{\vec{k}} = \vec{v}_{\vec{k}} + \frac{1}{V} \sum_{\vec{k'}}
f_{\vec{k},\vec{k'}} \, \delta(\mu-\epsilon_{\vec{k'}}) \,
\vec{v}_{\vec{k'}}.
\end{equation}

Notice that since we are transporting the whole Fermi sea by a
constant vector the change in the lagrangian must be
minus the total change in the energy of the system. Therefore,
$\vec{J}_{\vec{k}}$ is the contribution for current carried by a fermion
with momentum $\vec{k}$~\cite{baym}. Moreover, if the system is
homogenous and isotropic the current and the velocity must point
in the direction of $\vec{k}$. Naturally, the current transports
a mass equals to the mass of the fermion, $m$, namely,
\begin{equation}
\vec{J}_{\vec{k}} = \frac{\vec{k}}{m}.
\end{equation}
In the Landau theory the effective mass, $m^{*}$, is defined as the
coefficient of the velocity,
\begin{equation}
\vec{v}_{\vec{k}} = \frac{\vec{k}}{m^{*}}.
\end{equation}

Substituting (C5) and (C6) in (C4) and assuming $\vec{k}$ and $\vec{k'}$ at
the Fermi surface and $\vec{k}.\vec{k'} = p_F^2 \cos\theta$ we easily get,
\begin{equation}
\frac{m^{*}}{m} = 1 + \frac{1}{V} \sum_{\vec{k'}} f_{\vec{k},\vec{k'}}
\, \delta(\mu-\epsilon_{\vec{k'}}) \, \cos\theta
\end{equation}
which is the usual relation between the effective mass and the bare
mass in the Landau theory of the Fermi liquid~\cite{baym}.

In three dimensions we can expand the interactions in Legendre polynomials,
\begin{equation}
F(\theta) = \sum_{L} F_L P_L(\cos\theta)
\end{equation}
where $F = N(0) f$. Use the density of states as defined in (5.5)
and the orthogonality between the Legendre polynomials in order to
get,
\begin{equation}
\frac{m^{*}}{m} = 1 + \frac{F_1}{3}.
\end{equation}

In two dimensions we use (2.6) and the above definitions and we find,
\begin{equation}
\frac{m^{*}}{m} = 1 + \frac{F_1}{2}.
\end{equation}

The galilean invariance present in the argument of this appendix
shows that the mass which appears in the Fermi velocity in the case of a
homogenous system is not the bare mass
but the effective mass which is calculated in the above expressions.

\end{document}